\documentclass[final,authoryear,3p,times]{elsarticle}

    \usepackage{amssymb}
    
    \usepackage{booktabs}
    \usepackage{graphicx} 
    \usepackage{float} 
    \usepackage[none]{hyphenat}
    \usepackage{array}
    
    \journal{arXiv}

\begin{document}
    
    \begin{frontmatter}
    
    
    
    \title{Predicting Webpage Aesthetics with Heatmap Entropy}
    
    \author{
       Zhenyu Gu,
       Chenhao Jin,
       Zhanxun Dong,
       Danni Chang
    }
    \address{Interaction Design Lab, School of Media and Design, Shanghai Jiao Tong University, China}
    
    
    \begin{abstract}
    Today, eye trackers are extensively used in user interface evaluations. However, it's still hard to analyze and interpret eye tracking data from the aesthetic point of view. To find quantitative links between eye movements and aesthetic experience, we tracked 30 observers' initial landings for 40 web pages (each displayed for 3 seconds). The web pages were also rated based on the observers' subjective aesthetic judgments.
    Shannon entropy was introduced to analyze the eye-tracking data. The result shows that the heatmap entropy (visual attention entropy, VAE) is highly correlated with the observers' aesthetic judgements of the web pages. Its improved version, relative VAE (rVAE), has a more significant correlation with the perceived aesthetics. (r=-0.65, F= 26.84, P$<$0.0001). This single metric alone can distinguish between good- and bad-looking pages with an approximate 85\% accuracy. Further investigation reveals that the performance of both VAE and rVAE became stable after 1 second. The curves indicate that their performances could be better, if the tracking time was extended beyond 3 seconds.
    \end{abstract}
    
    
    \end{frontmatter}
    
    

    \section{Introduction}
\label{sec:intro}
Aesthetics is an important factor for engaging users in browsing a website.
Big data reveals that most visitors will determine whether to stay on or leave a webpage within just a few seconds after landing \citep{Liu2010}.
Appealing webpages give users a greater sense of trust
\citep{Casal2008The, Li2010Increasing, Lindgaard2011}.
Beautiful webpages are seemingly more usable \citep{Tractinsky2000}.

"Aesthetics" is rooted in the Greek term "aisthetikos", meaning "to perceive".
The human visual system, the eye and the connected cortex behind, enables us to perceive and recognize beauty around us.
An eye behaves like a spotlight \citep{Eriksen1972}: It has a narrow but very high-resolution foveal vision surrounded with a broader but much lower resolution peripheral area.
Visual attention keeps selecting where this spotlight moves toward and focuses upon, targeting things like faces, words, images on screens, and a variety of other meaningful objects, bringing salient details into focus and filtering out background clutter.
The spotlight keeps darting from one spot to another and our brain pieces all snapshots together to get a clear overall image of the scene.

Eye-tracking devices are designed for recording fixations (gaze) and saccades (eye movements).
In usability and marketing research, eye tracking has been used to evaluate the accessibility of key information.
Though useful, this is still a fairly limited measure of user experience \citep{Santella}.

Is it possible to use eye tracking to anticipate observers' aesthetic satisfaction with a webpage in the fashion of a mind reader?
This paper reports our efforts to answer this question.
The results are positive, giving evidence that eye-tracking devices do enable some new sense of beauty, as it where, in the eye of the beholder.

Thirty observers' visual explorations of 40 webpages were recorded individually using an eye tracker under normal laboratory conditions. The eye tracking duration for each page and each person was 3000 ms.
The webpages were later rated by the observers based on their personal aesthetic judgments.
Shannon entropy was introduced to analyze the eye tracking data.
The result shows that the entropy in heatmap (attention map), named visual attention entropy (VAE), is highly correlated with the observers' aesthetic judgements of the webpages. Its improved version, relative VAE (rVAE), has more significant correlation with the perceived aesthetics (r=-0,65; F= 26.84, P$<$0.0001).
This single metric alone can differentiate between good- or bad- looking web pages to a certain degree (appr. 85\%). Further investigation reveals that the performance of both VAE and rVAE became stable after 1 second. The curves indicate that their performances could be better if the tracking duration were longer than 3 seconds.
The entropy can be interpreted as an objective and the quantitative metric of chaos in our attentional processes.
Therefore, it could be direct evidence to the fluency hypothesis \citep{Reber2004}, which proposes that beauty is grounded in the processing experience of the perceiver.
The more fluently perceivers can process an object, the more positive their aesthetic response.

\section{Related Work}
\label{sec:rel}
\subsection{Eye tracking and experimental aesthetics}
It is a natural thought that our eye's attentional responses may have relations with our sensory contemplation or appreciation of an object.

To examine how eye movements relate to judging aesthetically pleasing works of art, \Citet{Berlyne1971} concluded that aesthetic evaluations are based on two types of visual exploration: one a global, diverse exploration and the other a local, specific information gathering.
The diverse exploration is characterized by long saccades and short fixation (gaze) durations, while specific information gathering has longer fixation durations and shorter saccades.
Berlyne proposed that the pattern of exploration, featured by the oscillation of the durations and saccade lengths, is crucial for judging images as aesthetically pleasing or not.
Berlyne's concept of visual exploration was influential to the following studies.
\Citet{Locher2006} demonstrated that simply altering the color-balance of the original abstract compositions will change the distribution of gaze and scan path.
\Citet{Franke2008} observed an increased number of eye fixations and longer durations on more preferable 3-D renderings.
\Citet{Plumhoff2009} observed that with pleasant graphics, fixation durations increase and saccade lengths fluctuate more over the viewing period.
\Citet{Wallraven2009}, analyzed eye-tracking data from 20 participants who looked at 275 artworks from different periods of styles, and found a strong effect of the art period (style) on both the number and duration of fixations.
\Citet{Massaro} used categorized (color, grey, human, and nature) art paintings as test material to investigate the contributions of bottom-up and top-down processes of visual attention to eye-scanning on the art paintings.
\Citet{Khalighy2015} developed an empirical aesthetic formula by conducting three eye-tracking experiments on simple geometric forms of stimuli.
He concluded that beauty is positively related to the product of the number of fixations and the standard deviation of the duration.

These aforementioned studies above revealed the potential of eye tracking in aesthetic studies.
The results so far, however, do not go significantly beyond Berlyne's initial work.
Some reported observations, such as the increased number of fixations, longer duration and oscillation of saccade lengths, etc., merely confirmed Berlyne's notion that eye activity on aesthetically pleasing objects is more active and dynamic.
It is difficult to interpret these results from an aesthetics point of view.
Although eye tracking has been increasingly used for aesthetics experiments, there is little research giving a convincing in-depth interpretation about how the eye's behavior connects with aesthetic responses.

\subsection{Eye tracking and webpage aesthetics}
Research on webpage aesthetics primarily focuses on finding predictable features within webpages such as visual complexity and order \citep{Deng2010}; low-level image statistics \citep{Zheng}; high level objective design metrics \citep {Ivory}. \Citet{Seckler2015Linking} examined how design factors such as structure and color are linked to different facets of subjective aesthetic perception of webpages.
\Citet{Reinecke} introduced computational models of perceived visual complexity and colorfulness of webpages, and they found the two models are predictive of people's aesthetic preferences.

Eye tracking has been extensively used in webpage evaluations for visualizing accessibility or interest bias toward specific objects.
Existing visualization and analysis tools have nothing to do with the user experience of aesthetics.
Proven quantitative links between eye movements and aesthetics are still lacking \citep{Santella}.
Can eye-tracking data provide a more general means of quantifying a viewer's aesthetic experience?
Answering this question may require some new perspectives to abstract and interpret the eye-tracking data.

\section{Hypothesis}
\label{sec:hyp}
Our idea is derived from the fluency hypothesis.
\Citet{Reber2004} proposed that beauty is grounded in a perceivers' fluency of processing an object.
Viewing a webpage is essentially a procedure of image processing in the human visual system.
How do we evaluate the fluency of this procedure?
Is there any observable behavioral difference between fluent and influent visual processes?
Eye movement, as the visible part of our attention, serves as a precursor to all other neurological and cognitive functions.
If the aesthetic pleasure is a function of the perceiver's processing dynamics \citep{Reber2004}, the attentional process must be the doorway to our aesthetic judgements.

At its most basic, attention is defined as the process by which we select a subset from all of the available information for further processing \citep{Eriksen1972}.
We propose that a fluent attentional process may have:
\begin{enumerate}
  \item Less conflict and more ease in competitive selection, which is the process determining which visual object gains access to working memory, between multiple attention cues.
  \item Less distraction and more ease in orienting and localizing attention onto an interesting object.
\end{enumerate}

To quantitatively measure the fluency of visual attention, this paper adopts entropy as the metric to the fluency/chaos in our visual attention behaviors and further explores whether it correlates with our aesthetic judgements. Several existing indices mentioned in section\ref{sec:rel} will also be investigated.

The Entropy measure will be applied on two aspects of eye-tracking data:
\begin{itemize}
  \item Entropy in gaze plot (Saccade sequence) represents the complexity and uncertainty in the temporal order of attentional shifts among multi-objects.
  \item Entropy in heatmap (Fixation distribution) represents the level of noise and de-concentration of spatial allocation of attention on the given objects.
\end{itemize}

\begin{figure*}
  \centering
  \includegraphics [width=0.8\columnwidth]{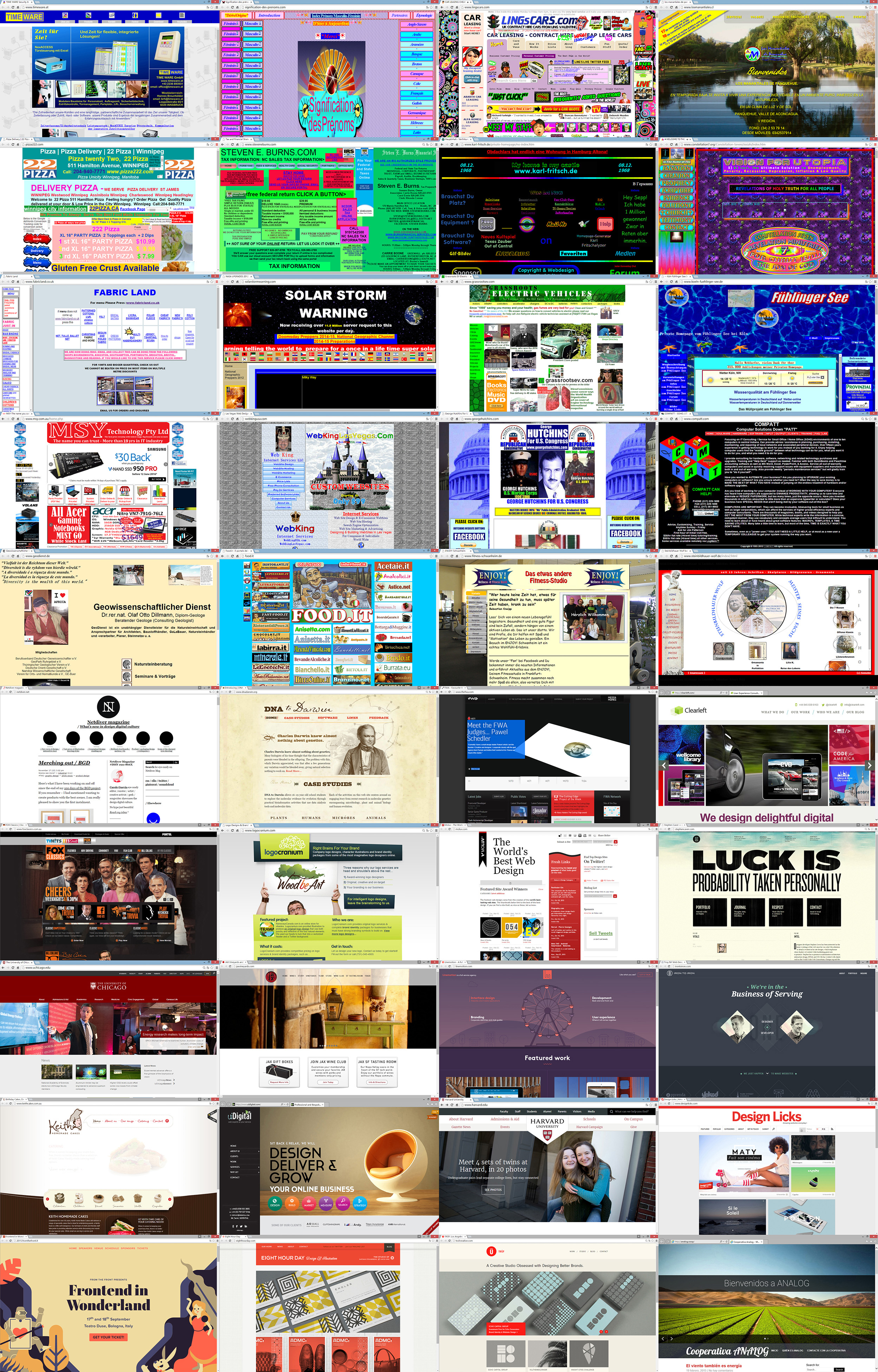}
  \caption{All the pages used in our eye-tracking test are arranged from top to bottom, left to right according to their scores from low to high.}
  \label{fig:all}
\end{figure*}

\clearpage
\section{Experiment and data collection}
\label{sec:exp}
The experiment simulates natural and accidental landings on websites. The only unnatural part of the experiment is the form we call Time-limited Exploration. That means the experiment encourages the observers to do quick general explorations on the pages. After three pages of dummy testing in advance, the observers should have been aware that they only had a few seconds to view each of the pages displayed. This is acceptable because we knew that users are used to making a quick decision whether staying or leaving in the first seconds after landing a new page to avoid wasting more time than necessary on bad pages \citep{Liu2010}.

The restricted exposure time could be enough to allow us to examine the early unintentional gazing behavior, and at the same time avoiding the fatigue of the observers.

\subsection{Equipment}
The eye tracker used was a Tobii T60 that works with the Tobii Studio software. The resolution of the eye tracker is 1280 pixels by 1024 pixel.

\subsection{Materials}
We selected snapshots of 40 websites' home pages as the testing material. To make the webpages more representative of high and low aesthetic qualities, we sampled pages from https://www.thebestdesigns.com and a site that collects votes for ugly web designs https://websitesfromhell.net/, and only selected the odd-numbered pages in the collections.

Pages that contained non-Latin characters or well-known logos and faces were removed to limit the evaluation to visual rather than socio-cultural aspects. Finally, we narrowed our selection to 20 webpages from each website. All pages were captured in landscape form at a resolution of 1280 pixels by 800 pixels to match the screen of the eye tracker.
The snapshot includes the frame of the web browser in order to better contextualize the experience to that of real web browsing.
Figure \ref{fig:all} shows all of the pages used as the stimuli.

\subsection{Subjects}
A total of 30 subjects (13 men, 17 women) participated in the eye tracking experiment. All were students from a number of departments in the university. Some of them were foreign students. Ages were between 19 and 27 years old.

\subsection{Site setting}
The experiment was conducted in a quiet room. The curtains of the room were pulled closed to avoid uncontrollable light and reflections.
The eye tracker was placed in front of a pure white wall to avoid possible distractions.

\subsection{Procedure}
The subjects were instructed to browse the webpages as though they were aimlessly wandering on the web. Each was seated in an office chair, and asked to lean forward to rest his/her chin comfortably on a soft, high support to fix his/her head 60cm from the eye-tracker screen. The subject's elbows rested on the table with a mouse in one hand, although any clicking and scrolling did not produce an interaction response.

In order to reduce fatigue, the test was divided into two phases, with a minute of rest in between. During each phase, the screen of the tracker automatically displayed half of the total of 40 experimental pages in a random sequence. Each page was presented for only 3 seconds, followed by 1 second of black screen. Three dummy pages were placed at the beginning of each phase to make the subjects feel comfortable with the rhythm of the test. During the eye tracking, there was no interaction between the operator and the subjects.

The subjective ratings of the pages were arranged separately after the experiment. Each subject was asked to review individually the experimental pages once again and judge the aesthetic quality of each page to be "good" or "bad". All subjects confirmed that they had never seen the pages before.

\subsection{Data collection}
\begin{figure}[H]
  \centering
  \includegraphics [width=0.9\columnwidth]{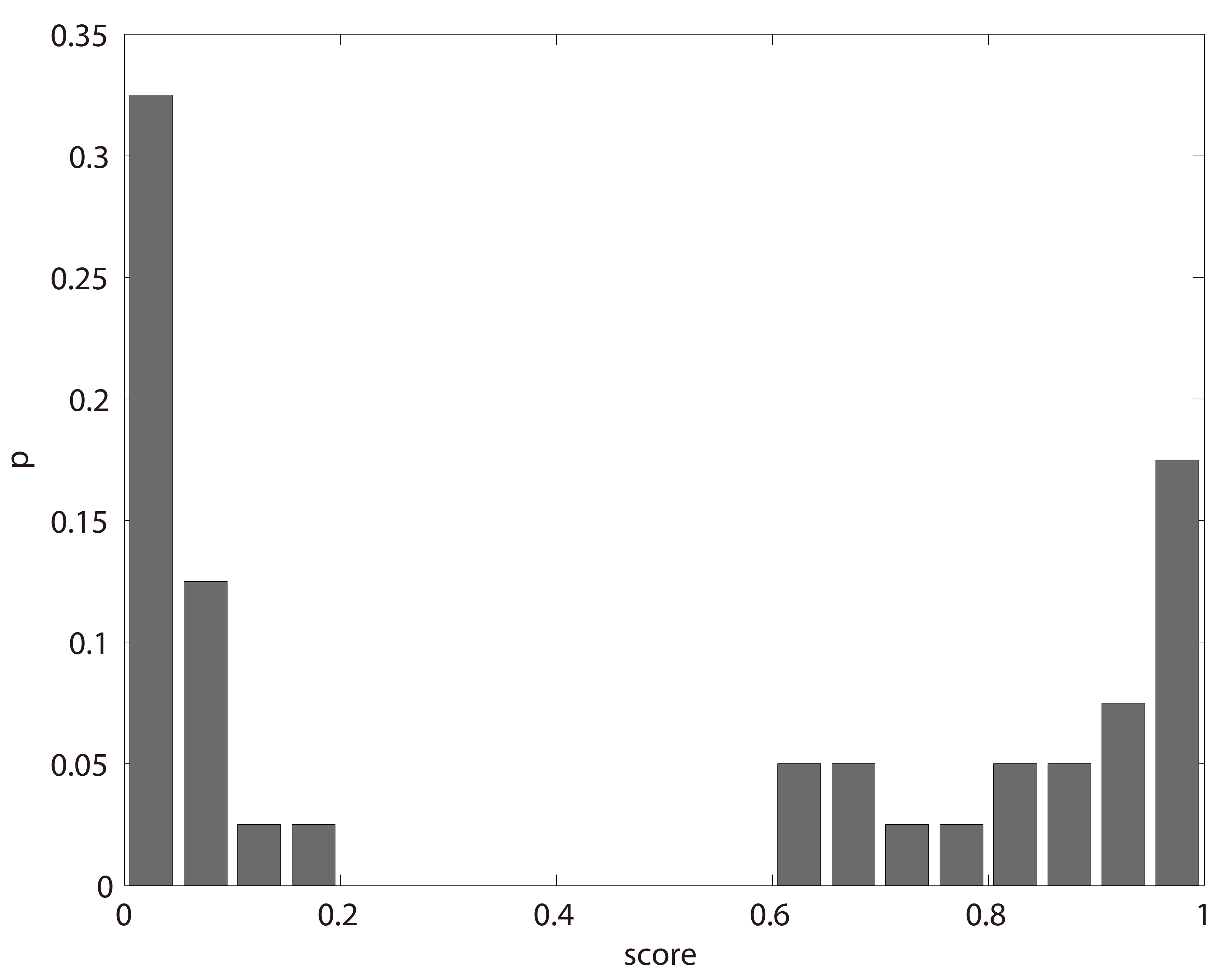}
  \caption{The distribution of the pages on scores. There are clearly two groups. The bad pages are generally more definite, while the good pages are more evenly distributed.}
  \label{fig:score}
\end{figure}

Each page in the test was given a final score based on how many "good" or "bad" scores it received. Figure \ref{fig:score} shows the distribution of the scores of the 40 pages.

The distribution is apparently bimodal. The 20 pages with scores less than 0.5 are exactly the same pages collected from https://websitesfromhell.net/. In other words, these 20 pages were recognized as "bad" pages without any doubt.

During operation, the Tobii T60 eye tracker has two main processes. In the first, it samples the eye positions on the screen at a frequency of 60Hz, and through interpolation it obtains the speed of eye movement on a constant basis. And then in the second process, it estimate where the fixation happens (if $speed=0$).

The raw data produced by the eye tracker consists of a series of fixations. Each fixation contains four parameters: the start time, the duration, and the X and Y position on the screen. The following analysis is just based on this format of eye-tracking data.

\section{Analysis}
\label{sec:ana}
\subsection{Traditional statistical indices}
We first applied some traditional descriptive statistics on the collected eye-tracking data. Most of these items have been used in the aesthetics researches described in section \ref{sec:rel}.

The indices we applied include:
\begin{itemize}
  \item number of fixations
  \item mean of duration
  \item standard deviation of duration
  \item mean of saccade length
  \item standard deviation of saccade length
  \item number of areas of interest(AOI)
  \item mean of AOI fixNum
  \item std of AOI fixNum
\end{itemize}

\begin{figure}[H]
  \centering
  \includegraphics [width=\columnwidth]{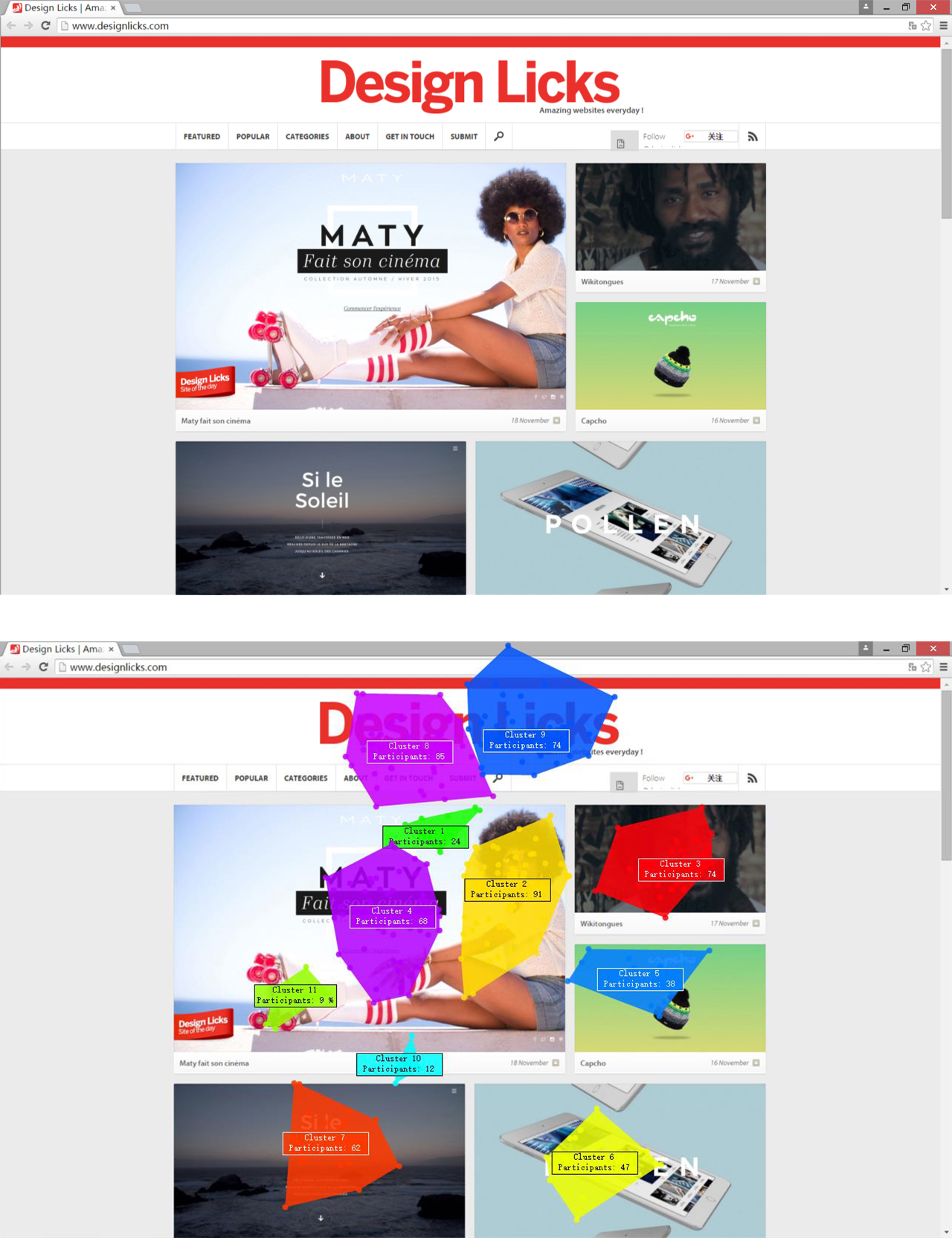}
  \caption{Top is one of the testing pages. Bottom is the page overlapped with AOIs, which are generated using default clustering tool in Tobii studio. There are total 11 AOIs for this page.}
  \label{fig:aoi}
\end{figure}

Areas of interest(AOI) are spatial clusters of the fixations on the pages. We used the default AOI clustering algorithm in the Tobii Studio package. Figure \ref{fig:aoi} is an AOI example obtained by the clustering.

The results of the analysis are listed in Table\ref{tab:traditional}. Each index has a linear correlation value with the aesthetic scores and a P-value of ANOVA test separating the "good" and "bad" classes of webpages.
The results show that only Number of fixations barely passed the ANOVA test (P$<$0.05).
All other indices failed the ANOVA test.
Number of fixations has a weak positive correlation with page aesthetics, more aesthetically pleasing pages generally got more gazes, implying the eyes were more active.
Meanwhile, number of AOIs has a weak negative correlation with the page aesthetics, meaning the aesthetically pleasing page's fixations are concentrated on fewer AOIs. It may also imply that subjects were more rapidly entering the stage of specific information gathering \citep{Berlyne1971} the variables of duration seem most irrelevant to the page aesthetics. It is not so strange considering the fact that the duration can range greatly by individuals from less than 100ms to well over a second \citep{ Irwin1996}.

\begin{table}[H]
  \centering
  \begin{tabular}{lrr}
     &correlation & ANOVA P-value \\
    \hline
    number of fixations & 0.3322 & 0.0429 \\
    mean of duration & -0.1526 & 0.3227 \\
    std of duration & 0.1165 & 0.6219 \\
    mean of saccade length & -0.2203 & 0.0871 \\
    std of saccade length & -0.1933 & 0.2732 \\
    number of AOIs & -0.2607 & 0.1631 \\
    mean of AOI fixNum & 0.3228 & 0.0782 \\
    std of AOI fixNum & 0.2556 & 0.0991 \\
    \\
  \end{tabular}
  \caption{The performances of some traditional statistics of fixations, saccades and AOIs. Only the number of fixations has a weak correlation with the aesthetic scores and the ANOVA is significant.}
  \label{tab:traditional}
\end{table}

\subsection{Entropies in eye movements}
Definitions of entropy reside in the fields of thermodynamics and information theory. Conceptually, entropy represents a system's chaos, degree of confusion, or vice versa, a lack of order and certainty.

The visual entropies defined in this paper are based on Shannon entropy, the entropy in information theory. Shannon entropy is a measure of the uncertainty based on the probabilities of all possible events.

The entropy $H$ (Greek capital letter eta) of a discrete random variable $X$ with possible values $\{x_1, ..., x_n\}$ and the probability mass function $P(X)$ is defined as
$$H(X)~=~-\sum_{i=1}^n P(x_i)\cdot log_{2}P(x_i)$$
The larger the $H(X)$, the higher the degree of chaos on the distribution of $X$, the greater the amount of uncertainty, whereas the smaller the $H(X)$, the more ordered the distribution of $X$ and the smaller the amount of uncertainty.

In the following, we analyze the visual entropies from two aspects of the eye-tracking data: saccades (gaze transitions) and gaze distribution (heatmap).

\subsubsection{Entropy in gaze transitions}
Entropy of eye movements was first introduced by \Citet{Tole1983} as a measure of cognitive loads, describing the changes of looking behaviors of eleven pilots in the conditions of different tasks. The model was also adopted in analysis of eye-tracking data for car driving \citep{Gilland2008}. \Citet{Hooge2013} used a similar metric called "Scan path entropy" to quantify gaze guidance to a specified target.

This paper follows Tole's definition of visual entropy based on Markov chain models.

The entropy represents the uncertainty of the transition between the fixations. As we knew, the number of possible positions of fixations is $1024\times800$, equal to the resolution of the images. That is a huge space of states. To estimate the probabilities of the transitions among those states, one would need a very large sample set which is impossible to get via eye-tracking tests. Tole's solution was to reduce the number of states by subdividing the view into a number of areas. In this research, we take advantage of the system-generated AOIs. Using this method, only the inter-AOI transitions need be taken into account. Inter-AOI transitions are relatively large saccades. A transition is thought to be an attention driven mechanism that requires a slight but unconscious attentional shift from the current target to peripheral target acquisition \citep{Henderson1993}. This is similar to the concept of global exploration suggested by \Citet{Berlyne1971}. By AOI clustering, we can convert each observer's eye movements on a page into a series of AOI jump sequences.(See Table \ref{tab:seq})

\begin{table}[H]
  \small
  \begin{tabular}{ll}
    1. & 6 - 7 - 11 - 3 - 11 - 10 - 9 - 2 - 3 - 4\\
    2. & 7 - 3 - 4 - 8 - 6 - 4 - 3 - 2\\
    3. & 3 - 5 - 9\\
    4. & 7 - 4 - 8 - 11 - 3 - 7 - 11\\
    5. & 7 - 4 - 8 - 11 - 4 - 7 - 2 - 9\\
    6. & 7 - 3 - 7 - 4 - 8 - 11 - 9 - 2 - 9\\
    7. & 7 - 8 - 4 - 7 - 11 - 2 - 9 - 2 - 3\\
    8. & 3 - 7 - 11 - 10 - 9 - 2 - 9 - 4\\
    9. & 7 - 11 - 2 - 6 - 7 - 8\\
    10.& 4 - 11 - 5 - 2\\
    11.& 7 - 3 - 11 - 4 - 8 - 7 - 2\\
    12.& 7 - 3 - 4 - 8 - 3 - 2 - 9\\
    13.& 6 - 3 - 4 - 8 - 3 - 10\\
    14.& 7 - 3 - 1 - 10 - 11 - 7\\
    15.& 7 - 11 - 8 - 4 - 7 - 8 - 9 - 2 - 6 - 2\\
    16.& 7 - 4 - 8 - 5 - 9 - 2 - 9\\
    17.& 11 - 7 - 11 - 10\\
    18.& 7 - 3 - 4 - 8 - 11 - 10 - 9 - 2\\
    19.& 7 - 11 - 10 - 2 - 9 - 4 - 8\\
    20.& 7 - 4 - 8 - 7 - 3 - 1 - 3\\
    21.& 7 - 11 - 10 - 2 - 9\\
    22.& 7 - 3 - 11 - 10 - 2\\
    23.& 3 - 4 - 1 - 3 - 8 - 4 - 11\\
    24.& 7 - 11 - 4 - 8 - 3 - 9 - 2 - 9\\
    25.& 7 - 4 - 8 - 3 - 2 - 9 - 8 - 4 - 11\\
    26.& 7 - 4 - 8 - 11 - 7 - 3 - 2 \\
    27.& 7 - 3 - 4 - 8 - 11 - 7\\
    28.& 6 - 7 - 8 - 4 - 7 - 8 - 11 - 4 - 2\\
    29.& 7 - 11 - 7 - 3 - 11 - 8 - 4 - 8 - 7\\
    30.& 3 - 7 - 4\\
    \\
  \end{tabular}
  \caption{Gaze transitions of the 30 subjects among the 11 AOIs in the image (see Figure \ref{fig:aoi}).}
  \label{tab:seq}
\end{table}

The gaze transitions are supposed Markov chains in nature. If a discrete time series $X_1,X_2,X_3,...$ satisfies $P(X_{n+1}=x~|~X_1=x_1,X_2=x_2,...,X_n=x_n)~=P(X_{n+1}=x~|~X_n=x_n)$, then the time series is called a discrete-time Markov chain. The Markov property of AOI sequences can be simply explained as "the AOI we glance toward is related only to the AOI we have just gazed upon".

The transitional probability information of a Markov chain can be summarized totally by a one-step transition probability matrix.
In the experiment, the matrix of probability (see table \ref{tab:mat}) is estimated by counting all the occurrences of the inter-AOI transitions in table \ref{tab:seq}.

\begin{table}[H]
\centering
\scriptsize
  \begin{tabular}{@{}lllllllllll@{}}
  0    & 0    & 0.67 & 0    & 0    & 0    & 0    & 0    & 0    & 0.33 & 0    \\
  0    & 0    & 0.14 & 0    & 0    & 0.14 & 0    & 0    & 0.71 & 0    & 0    \\
  0.08 & 0.16 & 0    & 0.28 & 0.04 & 0    & 0.16 & 0.04 & 0.04 & 0.04 & 0.16 \\
  0.04 & 0.04 & 0.04 & 0    & 0    & 0    & 0.15 & 0.63 & 0    & 0    & 0.11 \\
  0    & 0.33 & 0    & 0    & 0    & 0    & 0    & 0    & 0.67 & 0    & 0    \\
  0    & 0.17 & 0.17 & 0.17 & 0    & 0    & 0.5  & 0    & 0    & 0    & 0    \\
  0    & 0.05 & 0.3  & 0.22 & 0    & 0    & 0    & 0.14 & 0    & 0    & 0.3  \\
  0    & 0    & 0.17 & 0.26 & 0.04 & 0.04 & 0.13 & 0    & 0.04 & 0    & 0.3  \\
  0    & 0.73 & 0    & 0.2  & 0    & 0    & 0    & 0.1  & 0    & 0    & 0    \\
  0    & 0.43 & 0    & 0    & 0    & 0    & 0    & 0    & 0.43 & 0    & 0.14 \\
  0    & 0.1  & 0.1  & 0.17 & 0.04 & 0    & 0.21 & 0.08 & 0.04 & 0.29 & 0\\
  \\
  \end{tabular}
\caption{Markov transition probability matrix of the 11 AOIs.}
\label{tab:mat}
\end{table}

In the Matrix, $p_{ij}$ represents the probability of transition from AOI $i$ to AOI $j$.
Then a visual entropy based on the first-order Markov transition matrix is calculated by the formula:
$$H = \sum_{i=1}^n(P(i)\sum_{j\neq i} p_{ij}log_2(p_{ij}))$$
Where $p_{ij}$ represents the conditional probability of transition from AOI $i$ to AOI $j$. $P(i)$ represents the prior probability of $i$, that is, the probability of starting from AOI $i$, which is obtained by counting the frequency of occurrence of AOI $i$ in all AOI sequences on the statistics page.

$H_{max}$ is the maximum entropy of the current AOI transitions. This is obtained by assuming that all transition probabilities are equal and all the prior probabilities are equal.
By dividing the $H_{max}$, the visual entropies of different pages with different numbers of AOIs are comparable to each other.

$$H_{relative}~=~\frac{H}{H_{max}}$$

Given this method, entropy can inform us of the certainty of the attentional shifts between AOIs. We expected that the entropy of a good page would have a smaller value. However, the result was disappointing. Based on the 3s eye tracking data, The correlation between the entropies and the aesthetic scores was 0.1585, and the two classes' ANOVA P-value was 0.4741.

\begin{table}[H]
\centering
\begin{tabular}{lrrrrr}
  Source&SS&df&MS&F&Prob$>$F\\ \hline
  Groups&0.00166&1&0.00166&0.52&0.4741\\
  Error&0.12101&38&0.00318&&\\
  Total&0.12268&39&&&\\
\end{tabular}
\caption{ANOVA of visual entropy based on the gaze transitions.}
\label{tab:ANOVA-ve}
\end{table}

The failure of the entropy based on AOI Markov chain might have the following two reasons:
\begin{itemize}
  \item The first-order Markov chain might be an over-simplified model. However, increasing the order of the chain requires an exponentially larger data set.
  \item Relevant information was lost and some errors were introduced in the clustering and the default clustering algorithm in tobii studio was not smart enough to avoid incorrect clusters.
\end{itemize}

\subsubsection{Entropy in heatmap}
Entropy based on Markov chain regards eye movements as a serially-related events of eye fixations and saccades.

Here we introduce a new entropy metric based on the heatmap, which is the best-known visualization technique for eye tracking studies \citep{Nielsen2010}. In contrast to the gaze chain, heatmap has no information about the order of fixations. This means the new entropy metric assumes that all of the gazing events are independent.

\begin{figure}[H]
  \centering
  \includegraphics [width=\columnwidth]{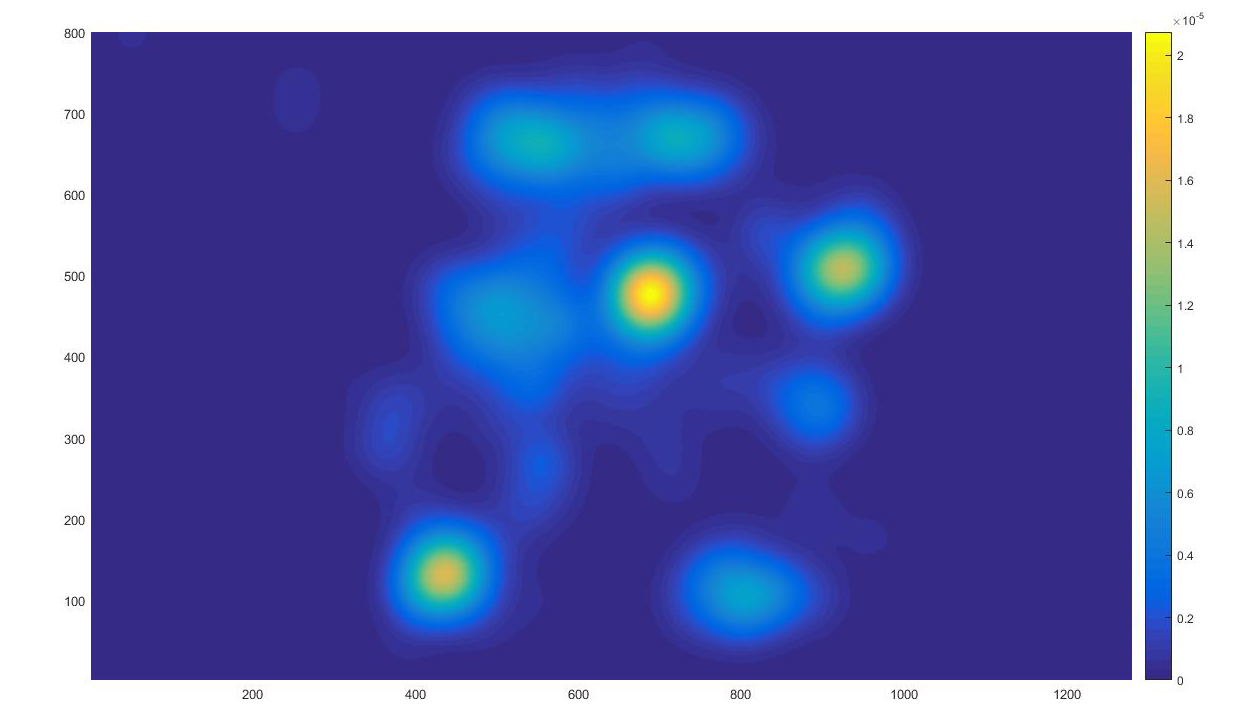}
  \caption{The heatmap of the example pages in figure \ref{fig:aoi}. It visualizes the spatial probability distribution of fixations on the page.}
  \label{fig:hm}
\end{figure}

Considering a two-dimensional random variable $(X, Y)$ which represents the position of a fixation, the resolution of the screen constitutes its probability space (i.e., each pixel is a possible position for a fixation).
Based on the probability distribution $P(X, Y)$, The visual entropy is defined as:
$$H(P)~=-\sum_{i=1}^{1280} \sum_{j=1}^{800} P(x_i, y_j)log_2(P(x_i, y_j))$$
where: $$\sum_{i=1}^{1280}\sum_{i=1}^{800}~P(x_i, y_j)~=~1$$

$P(X, Y)$, the probability distribution of $(X, Y)$, which is actually the heatmap, is posterior and can be estimated by the eye tracking data.

The entropy of the heatmap reflects the consistency or agglomeration of the subjects' eye movements in the space: when all the gazes concentrate on a single pixel, it yields the smallest value, and when the gazes are evenly scattered on the screen, it yields the maximum value.

To estimate the $P(X, Y)$ with the rather small sample of fixations, an important technique is Gaussian mixture, which interpolates all the fixations by placing a Gaussian kernel at each observed fixations.

For a single fixation $(x_0, y_0, d)$, where $x$ is the horizontal position, $y$ is the vertical position and $d$ is the duration,
The expression of the kernel is
$$d\cdot e^{-\frac{(x-x_0)^2 + (y-y_0)^2}{2\sigma^2}}$$
Where $\sigma$ is the standard deviation of the kernel.
Notice that the expression is not the standard normalized gauss distribution as the model will be normalized in the end to make sure the total probability $\sum\sum P(X,Y) = 1 $.

Placing a proper-sized Gaussian at a fixation is also reasonable given the fact that the fixation is actually a round area corresponding to the fovea. In addition, the fixation in reality is not absolutely fixed due to the fixational movements \citep{Martinez2004}. The mixture model also tolerates the unavoidable measurement error of the eye tracker. Figure \ref{fig:hm} is a typical heatmap with $\sigma=30px$.

The visual entropies of 40 pages based on their heatmaps turned out to be negatively correlated with their aesthetic scores. Pearson correlation r= -0.5412. ANOVA F = 15.79 P = 0.0003. The result is significant.

\begin{table}[H]
\centering
\begin{tabular}{lrrrrr}
  Source&SS&df&MS&F&Prob$>$F\\ \hline
  Groups&1.15861&1&1.15861&15.79&0.0003\\
  Error&2.78901&38&0.07339&&\\
  Total&3.94762&39&&&\\
\end{tabular}
\caption{ANOVA analysis of the entropy of heatmap.}
\label{tab:ANOVA-vae-dw}
\end{table}

Present results show that the visual entropy based on heatmap is the most promising index to predict a web page's aesthetic score. heatmap sometimes is also sometimes referred to as attention map. So we name this matric visual attention entropy (VAE).

\subsubsection{Relative VAE}
There is a defect with VAE as a metric. VAE is negatively correlated with aesthetics judgements. However, an absolutely lower VAE does not necessary mean a very high aesthetic quality. For instance, a web page with very little content may result in a very low VAE. On the other hand, a web page containing many objects may have relatively high VAE. This certainly does not mean that the former is better than the latter. Theoretically, the VAEs of the pages with different contents are not comparable to each other.

To solve this problem, we introduced a concept we call relative VAE (rVAE). To compare VAEs of different webpages, it is necessary to take into account their different base VAE (bVAE), which is supposed to be the noise-free VAE of a page. It represents the inevitable attention cost for the page. The bVAE on the page is estimated by averaging the individual VAEs of all subjects.

$$bVAE~=~\frac{1}{n}\sum_{i=1}^n VAE(P_i)$$

Where n is the number of subjects, $VAE(P_i)$ is the individual VAE measured on $i^{th}$ subject.
With the bVAE as a precondition, now we get the rVAE:

\begin{equation}
rVAE = \frac{VAE}{bVAE}
\label{formula:rvae}
\end{equation}

Compared to VAE, The performance of rVAE was significantly improved, the Pearson correlations increased from r= -0.54 to r= -0.66. ANOVA F=26.84 with P-value=0.000008.

\begin{table}[H]
\centering
\begin{tabular}{lrrrrr}
  Source&SS&df&MS&F&Prob$>$F\\ \hline
  Groups&0.00336&1&0.00336&26.84&7.53E-06\\
  Error&0.00475&38&0.00013&&\\
  Total&0.00811&39&&&\\
\end{tabular}
\caption{ANOVA of rVAE.}
\label{tab:ANOVA-rvae-dw}
\end{table}

The formula \ref{formula:rvae} indicates that the higher the bVAE, the lower the rVAE, therefore demonstrating the aesthetics. This is not fully true, however, and in reality the higher bVAE may result in possibly a much higher VAE, since the two indices are significantly correlated (r = 0.77, see Table \ref{tab:corr}). If bVAE is regarded as the signal in a channel of communication, rVAE minus 1 is actually the Noise-to-Signal ratio.

\begin{table}[H]
\centering
\begin{tabular}{l|rrrrrr}
        &score&fixNum&VAE&bVAE&rVAE\\ \hline
  score &1&0.33&\bfseries{-0.54}&-0.13&\bfseries{-0.66}\\
  fixNum&-&1&0.14&\bfseries{0.61}&-0.14\\
  VAE&-&-&1&\bfseries{0.77}&\bfseries{0.94}\\
  bVAE&-&-&-&1&0.52\\
  rVAE&-&-&-&-&1\\
\end{tabular}
\caption{The pearson correlations in-between the aesthetic scores and the main indices.}
\label{tab:corr}
\end{table}

\begin{figure}[H]
  \centering
  \includegraphics [width=1\columnwidth]{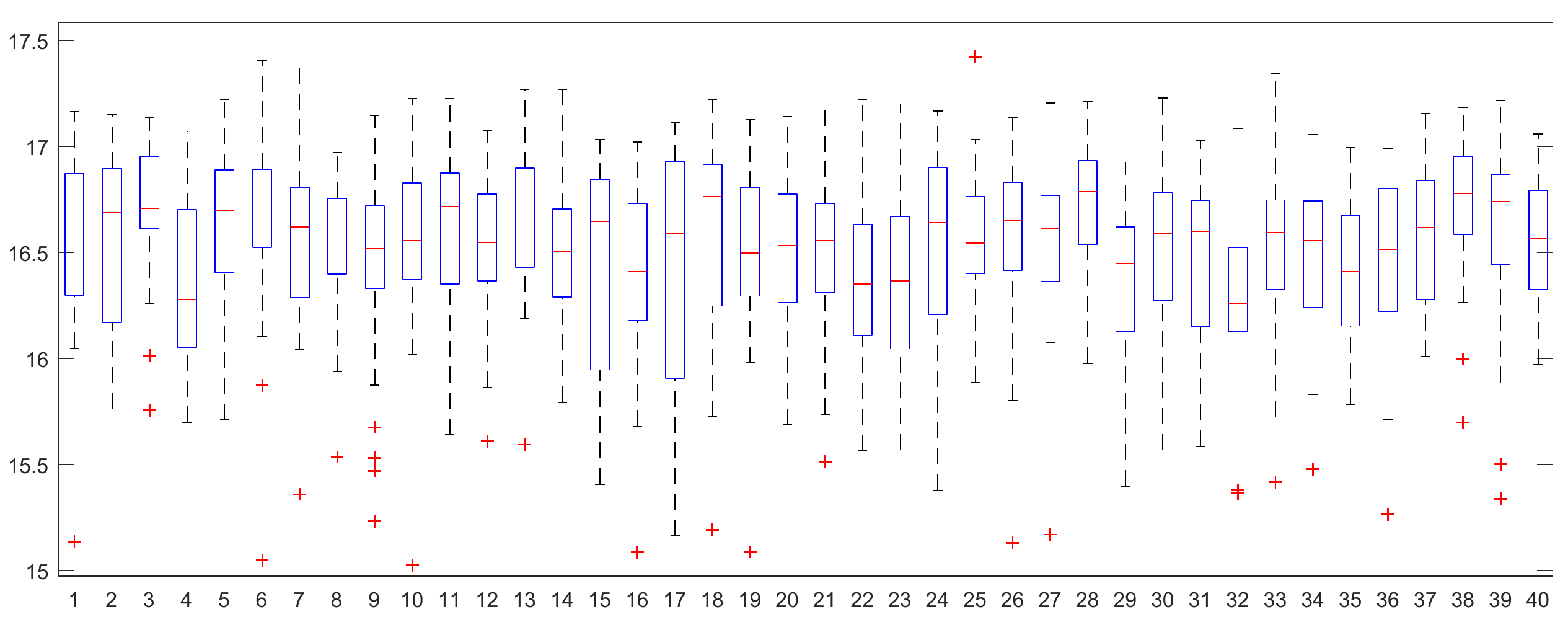}
  \caption{Boxplots of the 30 individual VAEs on each of the 40 webpages. The 40 pages are sorted by their scores, from low to high. The red lines in the boxes represent the bVAEs (means).  There is no apparent difference between the good and the bad pages}
  \label{fig:bvae}
\end{figure}

The analysis shows that aesthetic score is weakly correlated with Number of Fixation (r=0.33). So the good pages are generally have more fixations (interesting points). And bVAE is also positively correlated with Number of Fixation (r=0.61). This implies the good web pages generally are not based on lower bVAEs. Those facts may also explain why bVAE has ignorable linear correlation with the aesthetic score (r=-0.13). Figure \ref{fig:bvae} shows bVAE has no difference between the good and bad pages.

\begin{figure*}
  \centering
  \includegraphics [width=1.1\columnwidth]{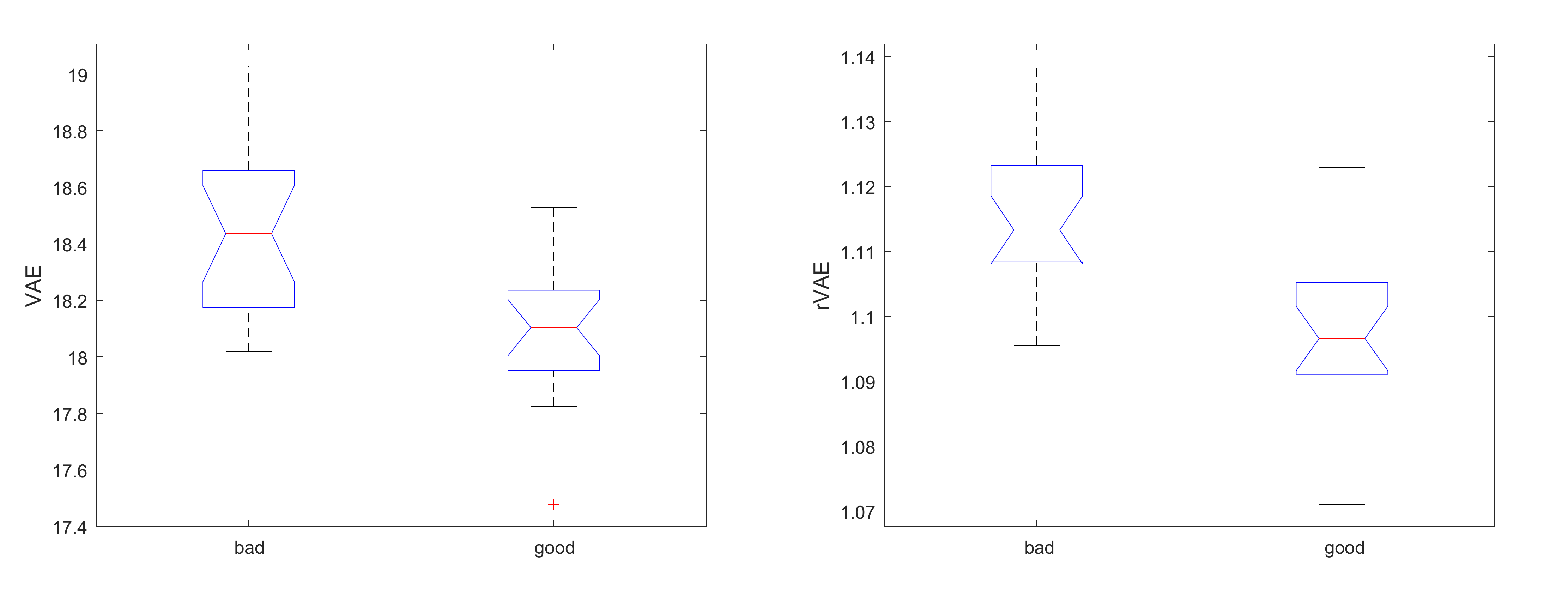}
  \caption{Boxplot of VAE (left) and rVAE (right). Apparently, the rVAEs more significantly separate the two classes.}
  \label{fig:box}
\end{figure*}

\begin{figure*}
  \centering
  \includegraphics [width=1.1\columnwidth]{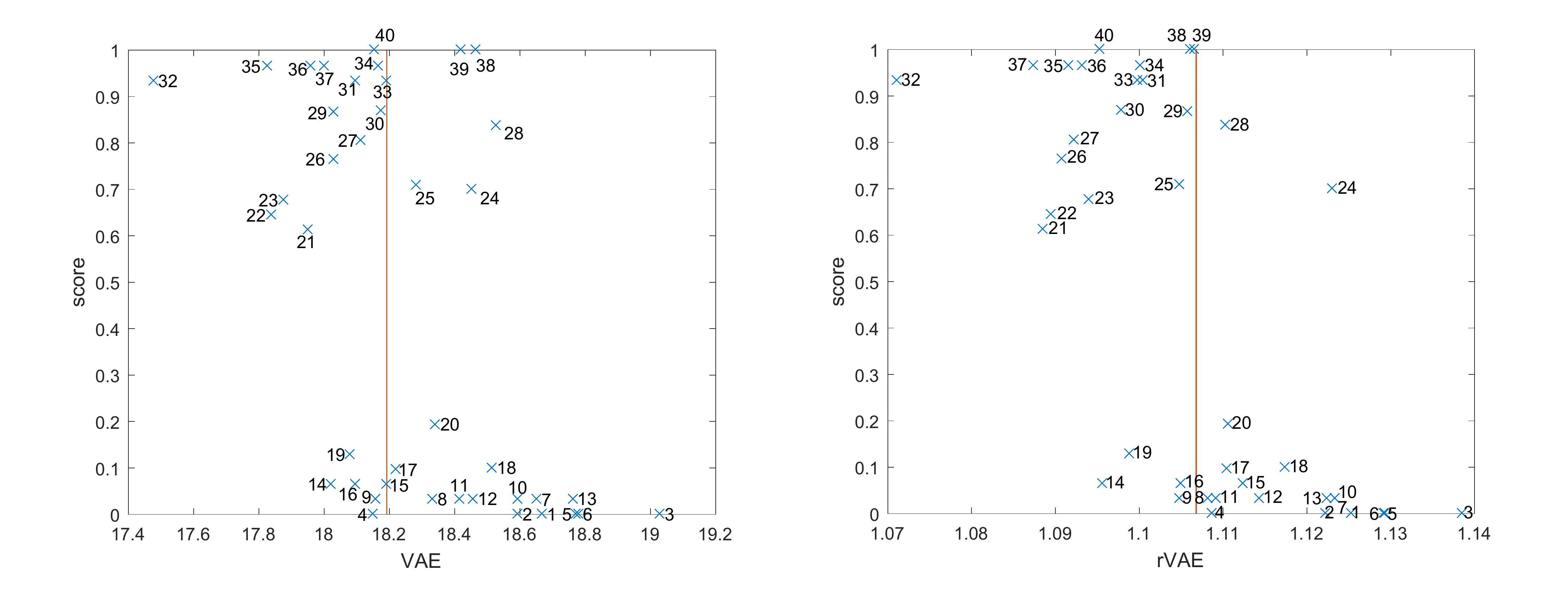}
  \caption{The scatter plots of 40 pages over VAE (left) and rVAE (right). The point cloud in the right plot appears more significantly skewed.}
  \label{fig:with-score}
\end{figure*}

\begin{figure*}
  \centering
  \includegraphics [width=\columnwidth]{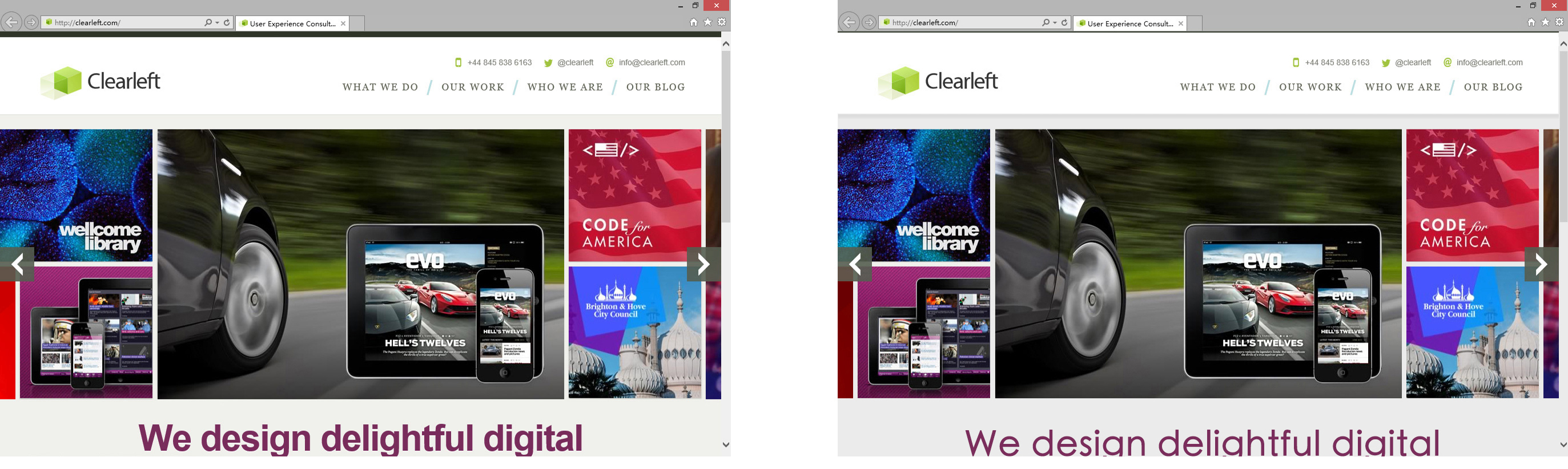}
  \caption{The left is one of the outlier pages (point 24 in figure \ref{fig:with-score}). If you display it on full screen, you may feel your eyes a little strained with selection pressure. As a comparison, the right is a revised version, in which some local saliencies have been adjusted. You may feel your eyes slightly relaxed.}
  \label{fig:out}
\end{figure*}

The introduction of bVAE enables us to compare VAEs of pages with different complexities of contents, and get more precise predictions to their aesthetics.

Figure \ref{fig:box} is the ANOVA boxplots of VAE and rVAE. Figure \ref{fig:with-score} is the scatter plots of 40 webpages based on their VAEs and rVAEs. Compared to the VAE, the rVAE performs better. The separation of high and low score pages are clearer.

\subsection{Outlier}

Comparing the two scatter plots in Figure \ref{fig:with-score}, we find that generally the points with higher scores (above 0.5) move slightly to the left, while the points with lower scores skew right. The rVAE metric makes them more separable. However, the metric alone is not able to fully explain the variation of the aesthetic scores. There are still some misclassified cases. An outlier (see the dot tagged 24) extremely deviated from the general tendency. Though it has an above-average score, its VAE and rVAE are quite high.

Checking the original image of the page (see Figure \ref{fig:out} left), we have to say that this is a not-bad-looking webpage with some appealing features, such as nice pictures, colors, symmetrical layout and so on. From a designer's point of view, however, it still has room for improvement. If you display it on full screen, you can feel your eyes a little strained with selection pressure. This may be attributed to some competitive visual saliencies in peripheral areas of the view. Figure \ref{fig:out} (right) is a slightly revised version of the page. It demonstrates how trivially adjusted local contrasts can make different overall impressions. Visual attention is essentially affected by bottom-up, low-level image features such as edge contrast or complexity.

So, VAE and rVAE is believed a universally influential factor of webpage aesthetics. The high VAE and rVAE in this case has exactly captured the unpleasant in observers' eyes, rather than an error.  Figure \ref{fig:out} right is a slightly revised version of the page. It demonstrates how trivially adjusted local contrasts can make a different overall impression.

\subsection{The stability of VAE}

\begin{figure}[H]
  \centering
  \includegraphics [width=0.85\columnwidth]{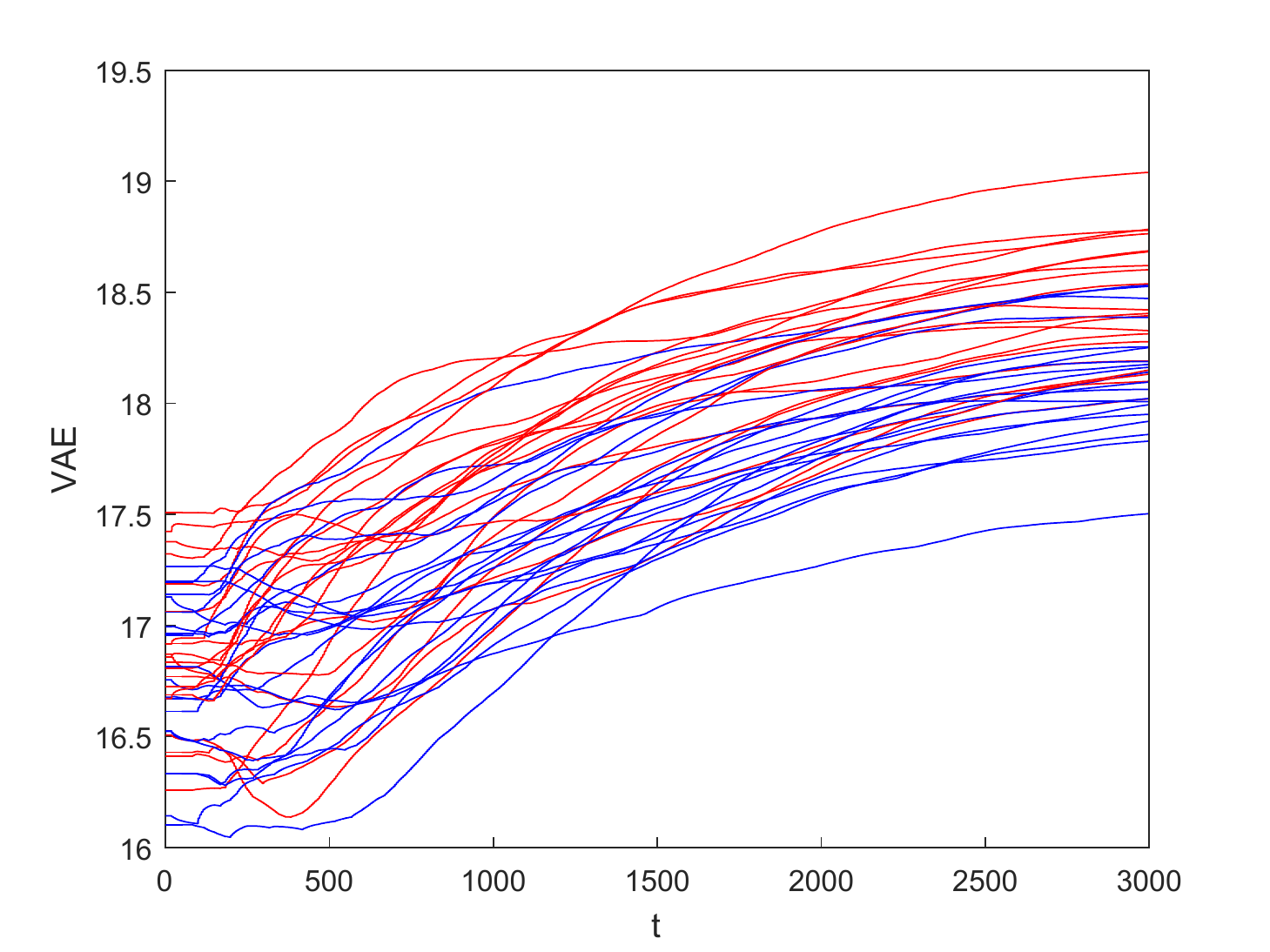}
  \includegraphics [width=0.85\columnwidth]{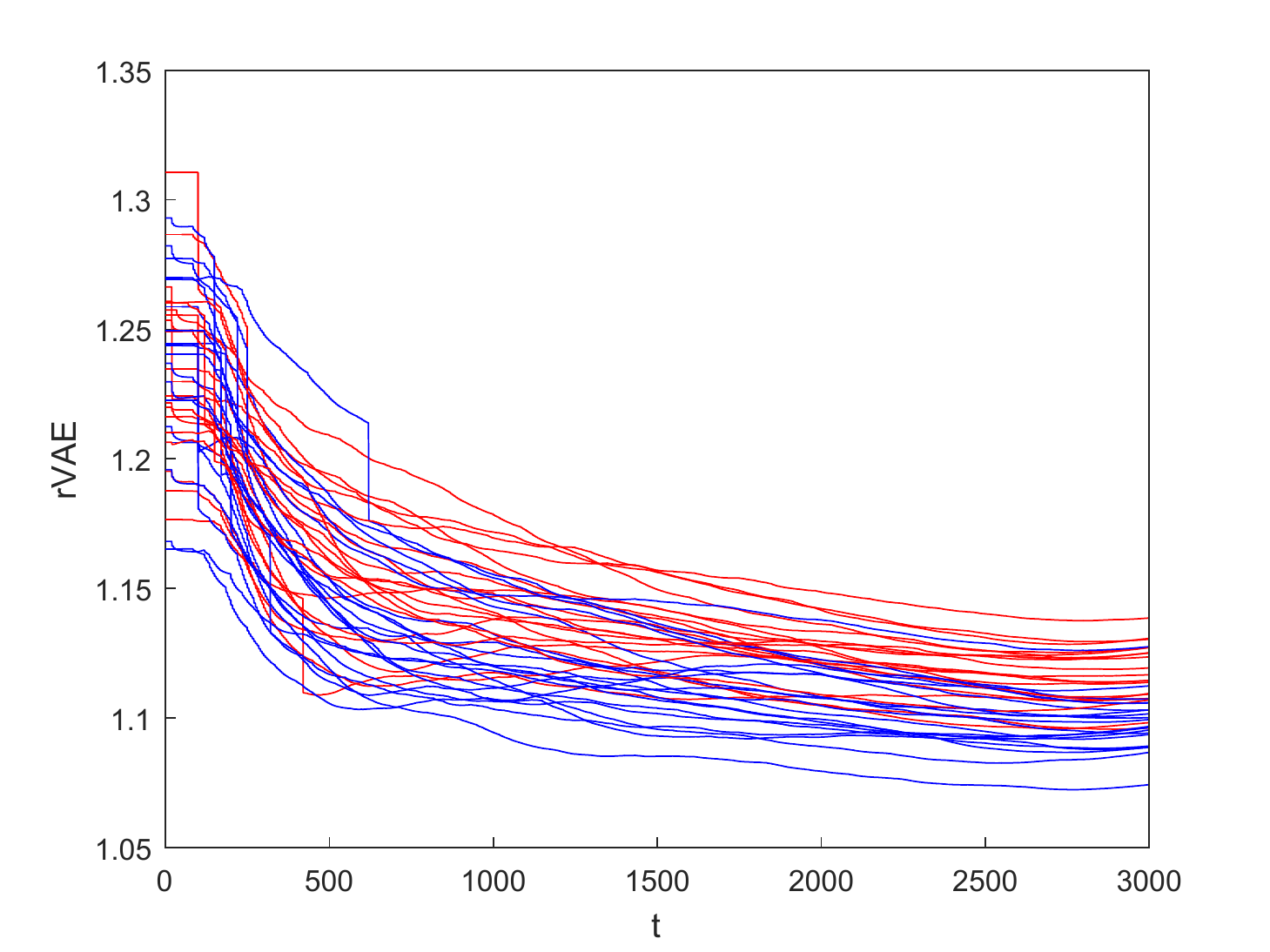}
  \caption{The VAEs (top) and the rVAEs (bottom) of the 40 pages gradually stabilize over time. The blue curves represent the "good" pages. The red curves represent the "bad" pages. The good pages generally yield lower VAE and rVAE values. The two groups  become more separated towards the end.
}
  \label{fig:with-t}
\end{figure}

For every moment of time t, we calculate the entropies based on the accumulated fixations during the period from 0 to t, until t = 3000ms,
obtaining a VAE-t curve and a rVAE-t curve for each of the 40 pages.
Figure \ref{fig:with-t} visualizes how the VAEs and rVAEs of the 40 pages change over time. All the curves fluctuate at the beginning, and then gradually flatten and stabilize.

\begin{figure}[H]
  \centering
  \includegraphics [width=0.85\columnwidth]{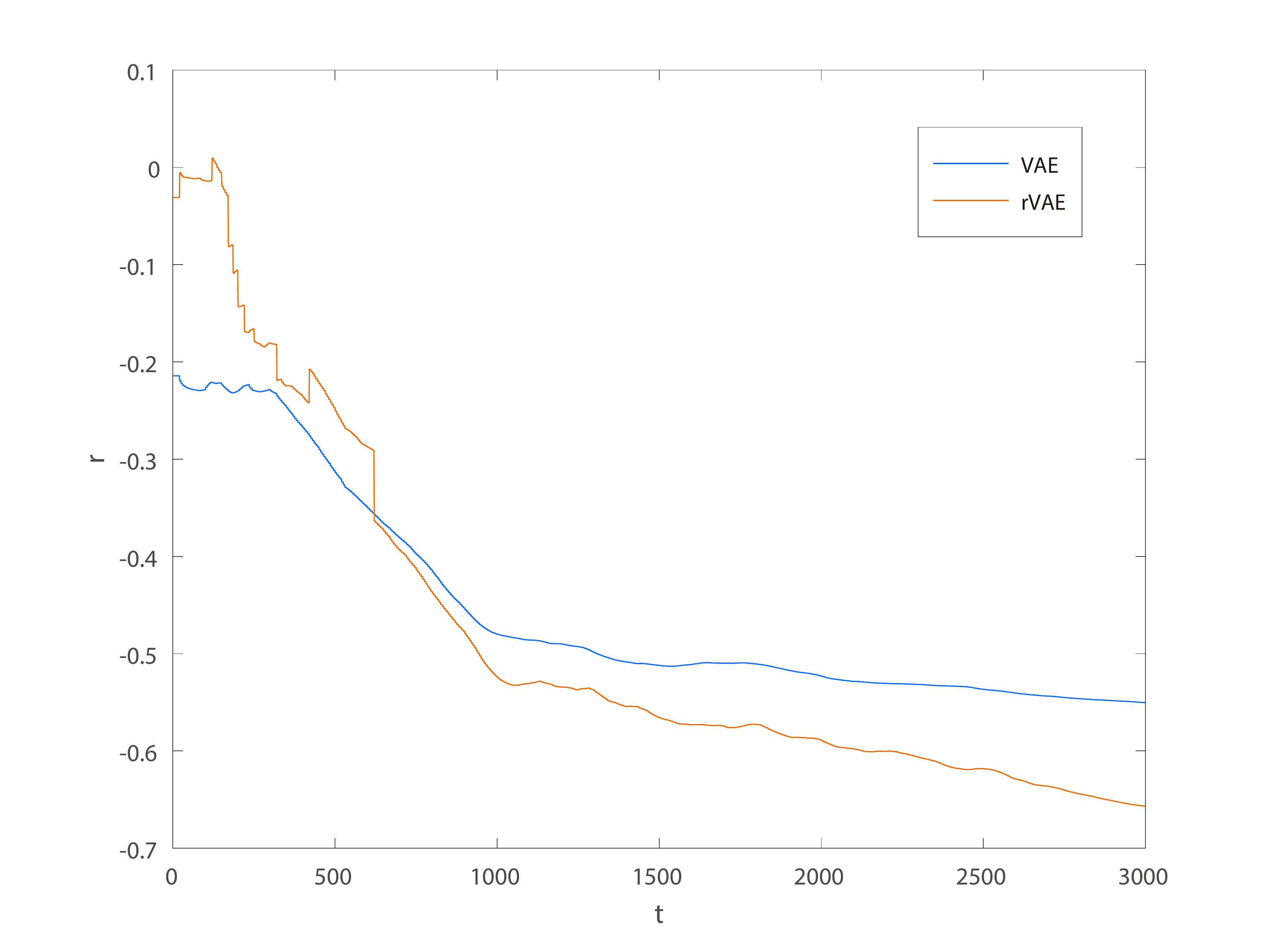}
  \caption{ The correlations between the entropies and the aesthetic scores become more and more significant over time. It is clear that rVAEs consistently sit below the VAEs after 1000 ms. The tendencies show that both curves, especially the rVAEs, could perform better, if the duration were longer than 3000 ms.}
  \label{fig:corr-t}
\end{figure}

In order to optimize the performance of VAE and rVAE, we further investigate how their correlations with the aesthetic scores vary with the testing time, the number of subjects and the $\sigma$ of the Gaussian kernel.
Figure \ref{fig:corr-t} visualizes how the correlations between the entropies and the aesthetics developing over the testing time. The correlations become most significant at 3s. As we did not record the eye movements after 3s, we are not able to see the development afterwards. We suppose the correlation can be stronger, judging from the tendency, which keeps decreasing.

\begin{figure}[H]
  \centering
  \includegraphics [width=0.85\columnwidth]{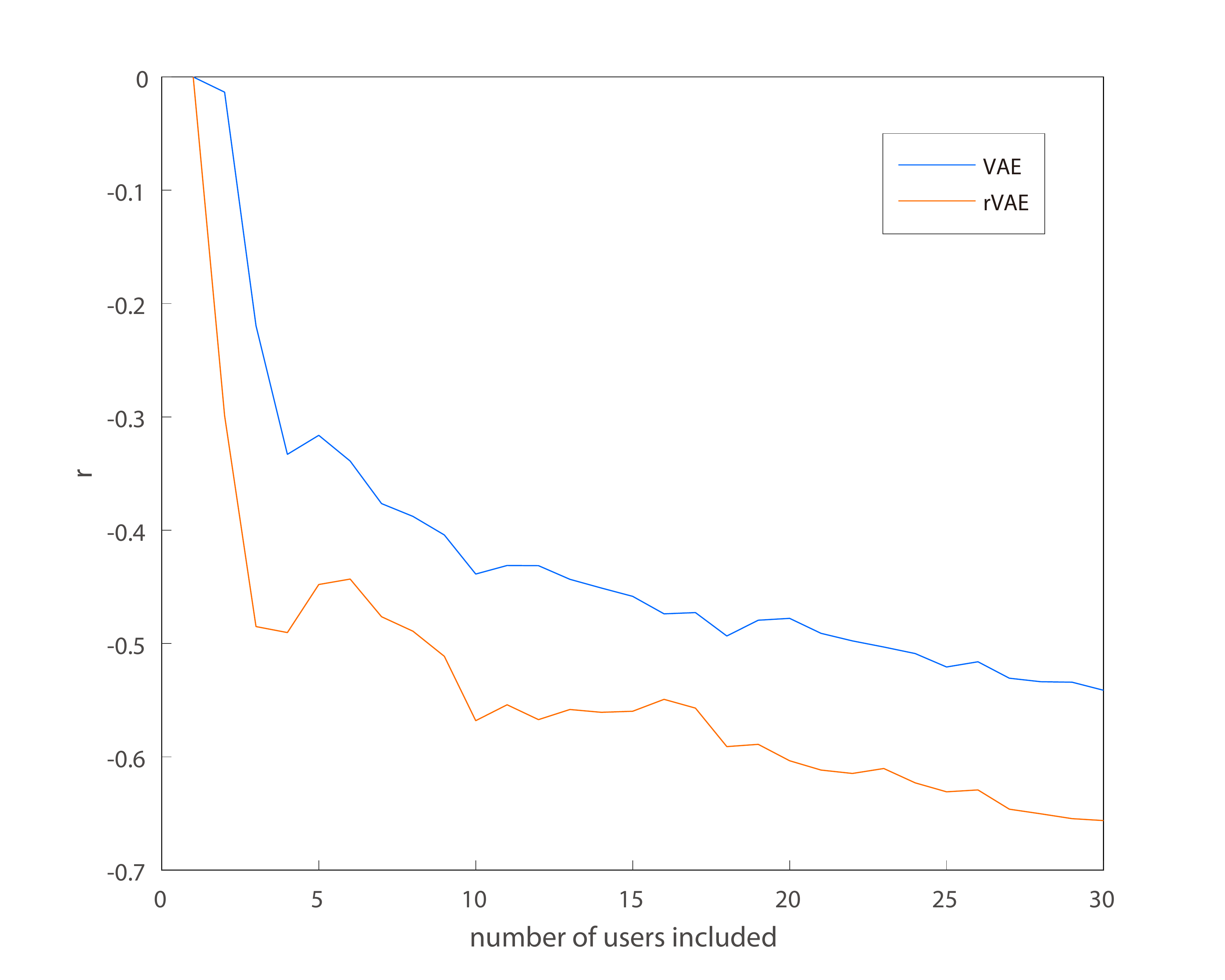}
  \caption{the correlations between the entropies and the aesthetic scores become more significant with the increasing numbers of subjects. The performance could be better if the number of subjects were larger than 30.}
  \label{fig:with-user}
\end{figure}

Figure \ref{fig:with-user} visualizes how the correlations between the entropies and the aesthetics become more significant with the increasing numbers of subjects. The sample sets of subjects are enlarged one by one, sized 2, 3, 4... 29, randomly picked from the total 30 subjects.

It seems the larger the number of subjects, the stronger the correlations. Adding more subjects is always welcome if without considering the limits of time and money.

Until now, all the analysis of the entropies is based on the heatmap with Gaussian $\sigma=30px$. How sensitive are their performances when the value of the $\sigma$ is changing?

\begin{figure}[H]
  \centering
  \includegraphics [width=0.85\columnwidth]{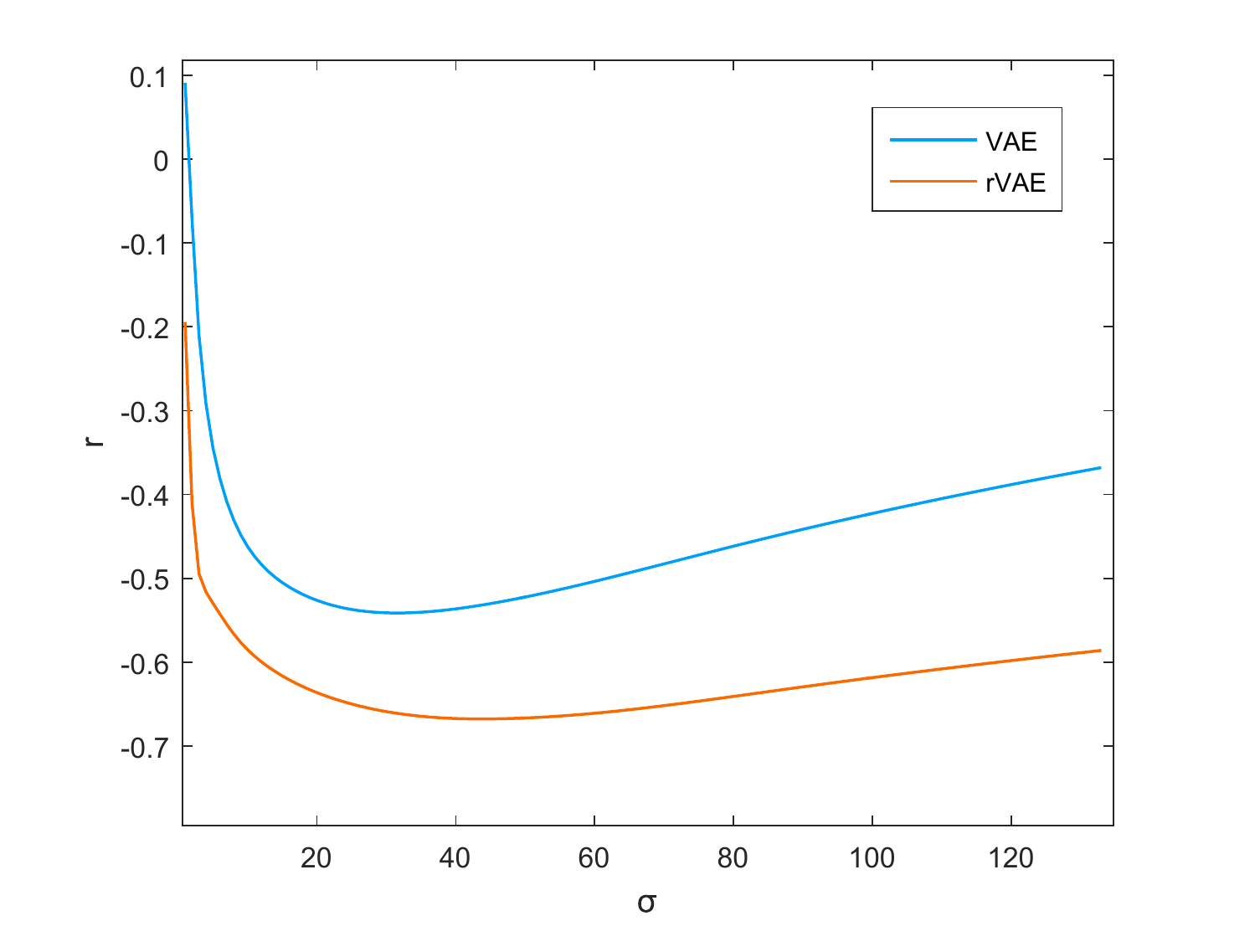}
  \caption{ The correlations smoothly change with the value of $\sigma$. The performances of both VAE and rVAE are stably approaching their maxima in a rather wide range of the $\sigma$. For VAE, from 20px to 60px, the correlation coefficient stays below -0.5. For rVAE, from 10px to 120px, the correlation coefficient stays below -0.6 and the curve reaches its extremum at about 40px.}
  \label{fig:with-sigma}
\end{figure}

Figure \ref{fig:with-sigma} visualizes how the correlations changing with the size of Gaussian. Both VAE and rVAE are not so sensitive to the variable of the $\sigma$. Their performances, especially the rVAE's, are stably approaching their maxima in a rather wide range of the $\sigma$.

The above analysis verified the stability of the VAE and rVAE in predicting our aesthetic judgements.
And evidently, their performances could be better, if with a longer tracking time, a larger number of subjects and a proper $\sigma$ value.

\section{Discussion and Conclusion}
\label{sec:dis}
The experiment demonstrated that our aesthetic judgments of webpages do relate to our eye movement behaviors. Shannon entropy measure was tested on two aspects of the eye tracking data: gaze sequence and heat map. The entropy in heat map, VAE, was shown to be predictive of aesthetic judgements of the subjects, while the entropy in gaze series (Markov chain) was not. VAE as the entropy of visual attention can be interpreted as the chaos in allocating limited attentional resource. Its improved version, rVAE, has significant correlation (r = -0.65) with our perceived aesthetics. This single metric alone can predict whether a webpage is aesthetically pleasing or not to a certain degree.

The success of VAE is at least partially attributable to interpolation with Gaussians, in the belief that the fixation distribution is continuous. And projecting the fixations during all the 3s on a single plane makes the data points denser. The failure of the entropy in gaze sequence is possibly due to the extra temporal dimension making the data points much sparser. It is difficult to conclude at this point that temporal information is irrelevant to our aesthetic judgements. Perhaps in future work a more informative model can be found to solve this part of the problem.

A lower VAE can be interpreted to mean that there is less effort in eye orientation, searching, and less distraction in viewing. In other word, it has lower perceptual or attentional load.
The evidence that aesthetically pleasing pages have lower VAEs supports the fluency hypothesis \citep{Reber2004}, and somewhat explains why "what is beautiful is usable" \citep{Tractinsky2000}. Or, "poor usability lowers ratings on perceived aesthetics" \citep{Tuch2012Is}.

VAE is based on specific information in eye fixations. As a new objective and quantitative index of eye tracking data, it is surely not merely a metric for aesthetics. VAE in nature are more related to our perceptual and motor-control ability. Visual gazing has functional similarity to mouse-pointing. This explains why some visual design principles for eye guidance are quite similar to Fitts' law \citep{MacKenzie1992}.

rVAE can be interpreted as using the lowest possible VAE to perceive as much useful information (noise free signal) as possible in the constraint of the display (bandwidth).  This explanation is in line with the aesthetic principle, "Maximum effect for minimum means" \citep{Hekkert2006}, which is rooted in evolutionary aesthetics \citep{Shimamura2012}. The theory argues that the basic aesthetic preferences of Homo sapiens have evolved in order to enhance survival and reproductive success. rVAE has potential to be used more generally in all the fields of visual design.

Eye movements represent the collaborations of overt and covert attentions. Though we found that to predict a web page's aesthetics requires at least 1000 ms of eye-tracking data, it has been reported that our first impression of a webpage is formed as quickly as within 50-500 ms \citep{Lindgaard2006}. 50 ms is barely enough for a snapshot, and our eyes are still at their original positions. It implies that brain can make aesthetic judgements immediately upon the covert attention map formed by the visual residual after a mere 50-ms exposure.

\section{Acknowledgements}

    \bibliographystyle{elsarticle-harv}
    \bibliography{main}
    
    
    
    
    
    
    
    \end{document}